\documentclass[10pt,twocolumn,revtex]{article}
\usepackage{graphicx}
\title{Long-Distance Diffusion of Excitons in Double Quantum Well Structures}
\author{Z. V\"or\"os,$^1$ R. Balili,$^1$ D.W. Snoke,$^1$ L. 
Pfeiffer,$^2$ and K. West$^2$\\
$^1$Department of Physics and Astronomy, University of Pittsburgh, 
Pittsburgh, PA 15260\\
$^2$Bell Labs, Lucent Technologies, Murray Hill, NJ }

\date{(\today)}
\begin{document}
\maketitle

  {\bf Abstract}.
In this Letter we report on lateral diffusion measurements of excitons at low temperature in double quantum wells of various widths. The structure is designed so 
that excitons live up to 30 $\mu$s and diffuse up to 500 $\mu$m. Particular attention is given to establishing that the transport occurs by exciton motion. The deduced exciton diffusion coefficients have a very strong well-width dependence, and obey the same power law as the diffusion coefficient for electrons.

\vspace{.5cm}

Excitons move in semiconductors and semiconductor structures, transporting energy from one point to another. Typical experiments with excitons in quantum wells do not observe motion of excitons over macroscopic distances, because the excitons have short 
lifetime due to close spatial wavefunction overlap and low diffusion constant due to 
disorder scattering.

In this Letter we report a system in which excitons move freely over distances of hundreds of microns. This opens up the possibility for both new applications and 
new fundamental studies. For example, exciton transport over long distances may be useful for energy collection. As shown previously \cite{apl}, excitons can be subjected to a drift force which moves them from one place to another. If the motion of excitons can be controlled in circuits over macroscopic distances, the energy collected over a large area can be directed to collection points, similar to the way biological photosynthesis is believed to occur.

Achievement of a truly delocalized gas of excitons also allows study of their fundamental behavior and phase transitions in an electron-hole system. For example, Bose-Einstein condensation of excitons has long been predicted \cite{bectheory}, but the theory typically assumes translational invariance, while in typical experiments with quantum wells, disorder plays a major role \cite{berman}.

Electron and exciton transport properties in quantum well systems have been the subject of a number of studies \cite{sakaki, heller, wolfeprb, hillmer}. From the theory of interface roughness scattering, it is expected that exciton mobility in narrow quantum wells will depend on the well width in the same way as electrons, but this has not been directly established until now. As is well known \cite{mendez,snoke}, double quantum well structures extend the lifetime of excitons. In our experiments the lifetime was extended up to 30 microseconds, and the diffusion length of the excitons up to 500 microns.  Thus in these systems the measurement of the diffusivity of the excitons and their effective mobility readily lend themselves to time-resolved optical imaging.

Our double quantum well samples were grown by means of molecular beam epitaxy (MBE) on $n$-doped GaAs (100) substrates with $p$-type capping layer. The GaAs wells were separated by a 40-\AA\ $\mathrm{Al}_{0.3}\mathrm{Ga}_{0.7}\mathrm{As}$ barrier and had widths of 80, 100, 120, and 140 \AA, respectively. Voltage was applied perpendicular
to the wells, causing the bands to tilt and the electrons and holes to separate, leading to the existence of ``indirect'' excitons, with electrons and holes in 
different wells, as opposed to ``direct'' excitons within a single quantum well. Both the the lifetime and the energy of excitons can be tuned by the electric field
\cite{snoke,szymanska}. We made an effort to reduce leakage current through the structure, since free carriers screen the Coulomb potential, reducing excitonic
lifetimes, and free carriers can scatter with the excitons, reducing the exciton diffusion constant. In order to do away with these unwanted effects, the outer barriers included superlattices built up of 20 successive layers of AlAs/AlGaAs, 
similar to, e.g., Ref. \cite{timofeev}. This reduced the dark current passing through the double quantum wells to below 1 $\mu$A/cm$^2$.

Electrons were excited into the conduction band by short (200 fs) laser pulses tuned to the direct exciton energy. The repetition period of the laser was 4 or 8 $\mu$s, so that the remainder of excitons from previous pulses was negligible. After the pulse, under the influence of the electric field, electrons and holes tunnel through the 40-\AA\ inner barrier into different wells, thus producing indirect excitons in less than 1 ns. The transport of indirect excitons was measured by imaging the sample onto the entrance slit of a spectrometer. The light intensity at the indirect excitonic wavelength as a function of both time and position 
is measured with an accuracy of 2 ns and 25 $\mu$m, respectively. The method is similar to that described in detail in Ref. \cite{gilliland}. To maintain a temperature of approximately 1.8 K, all measurements were conducted in a Janis Varitemp cryostat, and the samples were immersed in liquid He.

\begin{figure}
\hspace{-.3cm}
\includegraphics[width=0.45\textwidth]{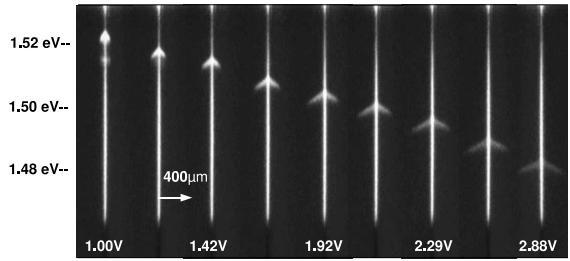}
\caption{Composite of the time-integrated luminescence from the 100-\AA \ well structure, recorded with an imaging spectrometer, for various applied voltages and with an excitation power of 2.7 mW. The narrow vertical line in the center of each image is impurity luminescence from the GaAs substrate, which is also excited by the laser. The chevron-like cloud spreading out from the central line shows the motion of the excitons away from the laser spot.}
  \label{fig:images}
\end{figure}

Fig.~\ref{fig:images} shows a composite of the time-integrated luminescence from the 100-\AA\ well structure, for various voltages, recorded by projecting an image of the sample onto the entrance slit of an imaging spectrometer. These images show both the 
spatial and the spectral profile of the luminescence. As the voltage is increased, the lifetime increases \protect\cite{szymanska}, and the excitons travel further.
In each of the images, there is a blue shift of the spectral position near the central excitation spot, which is the region of highest exciton density, and the spectral position falls as the density decreases far from the laser excitation spot. This shift comes from the repulsive interactions of the excitons, which lead to a mean-field energy shift, which is essentially just the Hartree term of the spatially seperated electrons and holes \cite{hartree}. As discussed below, this strong repulsion of the excitons gives a strong pressure-driven expansion of the exciton gas immediately after the laser pulse.

\begin{figure}
\includegraphics[width=0.45\textwidth]{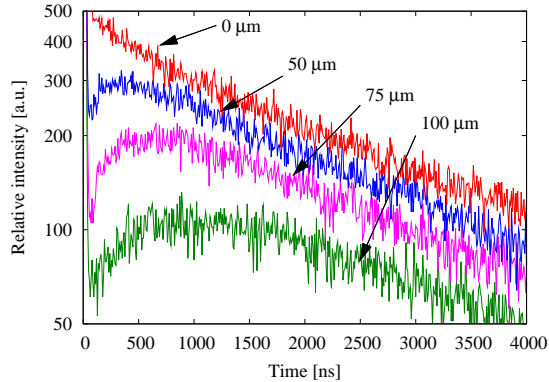}
\caption{Indirect exciton luminescence intensity as a function of time for various points on the sample with 100 \AA~ well width. The curves are labeled by the 
distance $x$ in the plane from the central laser spot. The average excitation power was 30 $\mu$W and the repetition period 4 $\mu$s.}
\label{fig:risetime}
\end{figure}

Fig.~\ref{fig:risetime} shows time-resolved luminescence for various distances from the
excitation spot. The rise time of the luminescence is longer for spots further away from the excitation spot, consistent with the picture that the excitons move out from the excitation spot without becoming localized. This figure also shows that the decay of the excitons at late times is single exponential; there is no evidence for any Auger
density-dependent recombination as seen for excitons in Cu$_2$O \cite{wolfeprb}.

Unlike previous experiments \cite{butov-nature,snoke-nature}, there is no dark region between the laser excitation spot and the exciton luminescence at long distance; the exciton cloud moves continuously from the laser spot outward. As shown earlier \cite{ssc,prl}, the effect of the ring of luminescence around a dark region came about because of a balance of carriers tunneling through the barriers and excitation of carriers over the barriers. In the present experiments, as discussed above, we have designed the sample to greatly decrease the tunneling through the outer barriers, and the laser excites the excitons resonantly, so that hopping over the barriers is suppressed. In previous experiments \cite{ssc}, no ring and accompanying dark region was seen when the laser photon energy was less than the Al$_x$Ga$_{1-x}$As barrier height.

\begin{figure}[!h]
\includegraphics[width=0.45\textwidth]{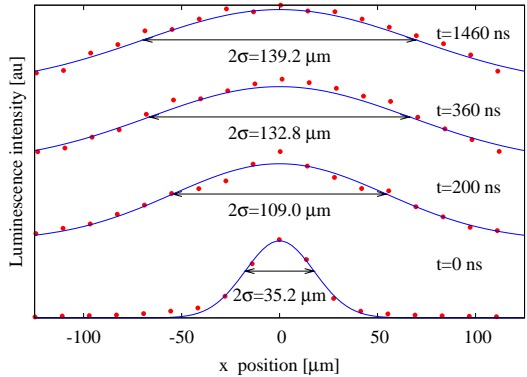}
\caption{Expansion of the exciton cloud at different times after the excitation. In this particular case the well-width was 100 \AA. The measured luminescence intensities are normalized. Also shown are the Gaussian fits with the value of the variance. The conditions are the same as in Fig.~\ref{fig:risetime}.}
  \label{fig:expansion}
\end{figure}

We can quantitatively measure the diffusion constant of the excitons using the
time-resolved luminescence. Fig.~\ref{fig:expansion} shows a series of spatially and
temporally resolved exciton emission profiles. In this case the width of the wells
was 100 \AA. The data in the figure are fit to Gaussian distributions. The widths of these fits are then used to get an estimate for the diffusion coefficient. As is 
well known, in the case of a purely diffusive process, the square of the Gaussian 
variance increases linearly with time. Keeping in mind that the solution of the 2D cylindrically symmetric diffusion equation is 
\begin{equation}
n(r, t) = \frac{n_0}{t} \exp\left(-\frac{r^2}{4Dt}-\frac{t}{\tau}\right) \ ,
\end{equation}
we get that the slope of the linear regime is twice the diffusion coefficient. In the equation above, $r^2 = x^2 + y^2$, $n(r, t)$ is the density of excitons, $n_0$ the initial density, $D$  the diffusion constant and $\tau$ the lifetime. Note that the finite lifetime of excitons causes an overall drop in the intensity with increasing time, but has no effect on the variance of the Guassian fit.

\begin{figure}[!h]
\includegraphics[width=0.45\textwidth]{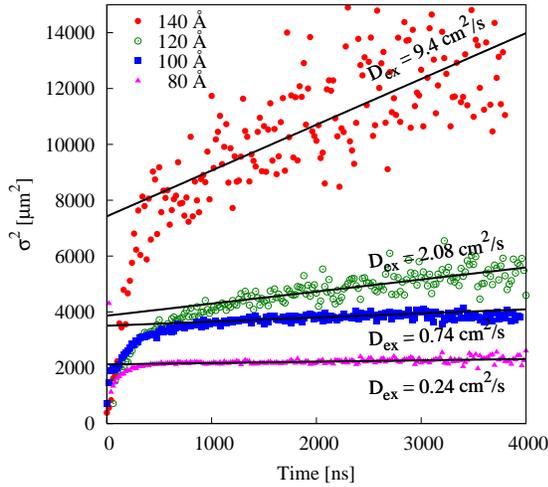}
  \caption{Measured variance-squared vs. time for the four samples, for average laser power 30 $\mu$W and pulse period 4 $\mu$s. The straight lines represent linear fits to the measured values in the range of 1000-4000 ns, i.e., when the expansions is purely diffusive.}
  \label{fig:variance}
\end{figure}

A typical plot of the time-dependence of the variance-squared for all four double well samples is shown in Fig.~\ref{fig:variance}. Immediately after the laser pulse the exciton cloud expands rapidly, then at late times, the behavior becomes linear. The initial period can be explained if one takes into account that following the pulse 
the exciton density is very high. This results in a pressure due to dipole-dipole repulsion between aligned excitons.  This pressure causes drift-like motion of the the 
excitons to move away from the excitation spot. A secondary effect of exciton density can be the filling up of local minima by excitons \cite{ivanov}, which leads to higher diffusion coefficient. At late times, after the density has dropped, pure excitons diffusion sets in and starts to dominate. 

\begin{figure}[!h]
\includegraphics[width=0.45\textwidth]{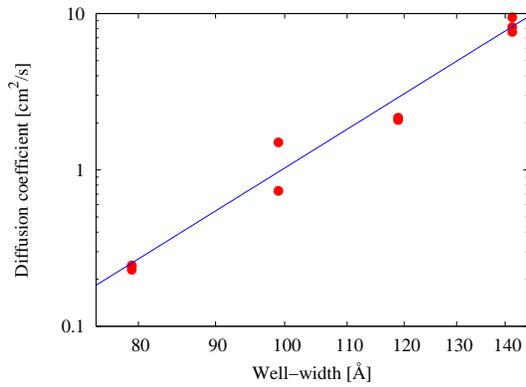}
  \caption{Measured diffusion coefficient as a function of well-width. 
For comparison, also shown is the theoretical dependence, $D \propto L^6$.}
  \label{fig:diffcoeff}
\end{figure}

Fig.~\ref{fig:diffcoeff}  shows the diffusion coefficients for the four double well samples.  The excitation density in all cases was kept low, in order to minimize the time in which the expansion of the exciton cloud is nondiffusive due to high dipole-dipole pressure.  As seen in this graph, there is an overall increase in the diffusion coefficient as the well width is increased, consistent with a $D\propto L^{6}$ power law. This is in accordance with both previous measurements for electron
transport \cite{sakaki} and a straightforward theory.

At these low temperatures, there are two main contributors to the diffusion coefficient, both increasing with well width. The first one is exciton-phonon scattering, whose well-width dependence is given by $\tau_{ph} = L/AT$, where $T$ is temperature, $A$ is a constant describing the exciton-phonon interaction strength and $L$ is the well-width \cite{basu, oberhauser}. The second effect is interface roughness scattering, whose well-width dependence is given by $\tau_{IRS} \propto L^{6}$ in the case of infinitely high barriers, assuming Rayleigh scattering from well-width variations which have relative potential energy difference $dE\propto dL/L^3$.  The power law determined here is in remarkably good agreement with the result of Sakaki et al. \cite{sakaki} for free electrons. This is consistent with the view that at in the low-temperature, dilute limit, all quasiparticle scattering from interface roughness is dominated by Rayleigh scattering.

In summary, in our double quantum well samples the lifetime of excitons can be extended up to 30 $\mu$s, thus making transport measurements feasible by means of
conventional optical techniques. The interactions of the excitons lead to a fast, pressure-driven expansion immediately after the laser pulse creates the excitons. At late times, the exciton motion is diffusive, and the diffusion coefficient has dependence on the well width consistent with a universal power law which applies when the scattering is dominated by Rayleigh scattering from well-width variations.

Although long-range diffusion of excitons has been demonstrated for twenty years in bulk semiconductors \cite{wolfeprl}, similar behavior has not been seen in quantum well
structures. This study shows that it is proper to treat the excitons as a gas of diffusing particles, and opens up the possibility of controlling the motion of excitons in chips over long distances.

{\bf Acknowledgements}. This work has been supported by the National Science Foundation under grant DMR-0404912 and by the Departent of Energy under grant DE-GF02-99ER45780.

\end{document}